\documentclass{ws-ijmpd}
\usepackage{t1enc}
\usepackage[latin2]{inputenc}
\usepackage[english]{babel}
\usepackage{amsmath, amsfonts, amssymb, mathrsfs}
\usepackage{graphicx, float}
\usepackage[toc,page]{appendix}
\usepackage{verbatim}
\usepackage{cite}
\usepackage{url}
\usepackage{subfiles}
\usepackage{xcolor}

\usepackage{hyperref}
\hypersetup{
    bookmarks=true,         
    unicode=true,          
    pdftoolbar=true,        
    pdfmenubar=true,        
    pdffitwindow=false,     
    pdfstartview={FitH},    
    pdftitle={4PN analysis of OJ287},    
    pdfauthor={Balázs Kacskovics, Mátyás Vasúth},     
    pdfsubject={Gravitational waves},   
    pdfcreator={Kacskovics Balázs},   
    pdfproducer={Dr. Vasúth Mátyás}, 
    pdfnewwindow=true,      
    colorlinks=true,       
    linkcolor=red,          
    citecolor=green,        
    filecolor=magenta,      
    urlcolor=cyan           
}

\begin{document}
	\title{Examination of orbital evolution and gravitational waves of OJ 287 in the 4PN order}
	\author{
		Bal\'{a}zs Kacskovics$^{1,2}$\thanks{email: kocskovics.balazs@wigner.hu}\, and M\'{a}ty\'{a}s Vas\'{u}th$^{2}$\thanks{email: vasuth.matyas@wigner.hu}
		}
	\address{ ${}^{1}$University of Pécs \\
 			H-7624 Pécs, Ifjúság Str. 6. \\
 			${}^{2}$WIGNER RCP, RMKI \\
 			H-1525 Budapest, P.O. Box 49.
		}

\maketitle

\begin{history}
	\received{Day Month Year}
	\revised{Day Month Year}
\end{history}

\begin{abstract}
We computed the orbital evolution and the emitted gravitational radiation of the supermassive black hole binary OJ 287. Here we used the initial data provided by the outburst structure of the system. We considered the spin-spin, spin-orbit, and the next-to-next leading order fourth post-Newtonian (4PN) corrections in our analysis. In this way, we could make an accurate examination of unstable orbits ($3M < r <6M$) of the secondary black hole. We tested the 4PN terms by analyzing the total and radiated energy, compared the post-Newtonian parameters and the separation of the two black holes for 3PN and 4PN. In conclusion, the 4PN corrections provided a significantly more accurate tool for analyzing unstable orbits than earlier 3PN terms. Furthermore, in this paper, we show the gravitational waves emitted by OJ 287 during its complete orbital evolution including unstable orbits.
\end{abstract}

	\section{Introduction}
	The supermassive black hole (SMBH) binary system OJ 287 is one of the few most frequently observed astrophysical objects, which presents us with an excellent test subject for General Relativity. Owning to the fact that all of its initial parameters are given to high accuracy by \cite{Valtonen2010, Valtonen2010a}, we can analyze its entire orbital evolution. OJ 287 is located at redshift $z = 0.306$, and the masses of the primary and auxiliary SMBHs are $m_1 = 1.84 \times 10^{10}~{\rm M_\odot}$ and $m_2 = 1.46 \times 10^{8}~{\rm M_\odot}$ which makes the components of this system one of the most massive objects in the seeable universe. For centuries, astronomers have observed and examined its quasi-periodic light variations, yet it has only recently been recognized as a blazar. The optical light curve of OJ 287 of the last $\sim 120$ year has been showing a double periodicity of 60 yrs and 12 yrs \cite{Valtonen2006, Valtonen2014}. In 1996 a detailed model was introduced by M. J. Valtonen and H. J. Letho \cite{Lehto1996, Sundelius1997}. In the model, the proposed 12-yrs orbital period is related to the major outbursts of the primary SMBH. The mechanism that forms the pair of outbursts occurring, 1 to 2 years separated, was assumed to be caused by impacts on the accretion disk of the primary SMBH by the secondary. During the transition of the auxiliary SMBH, matter from the accretion disk is taken away, and bremsstrahlung bursts are formed as the hot plasma becomes optically thin. This outburst structure allows us to reproduce the orbits of the binary with high accuracy. A more detailed analysis of the outburst structure can be found in \cite{Valtonen2010, Valtonen2010a}. From the luminosity of the source, we can determine both the masses and the relative velocity. It turned out that this system has a very high velocity of $v/c \simeq (0.06 \sim 0.26)$, and therefore it is an ideal candidate to test strong gravitational effects.

	From the timing of outbursts, Valtonen and his associates \cite{Valtonen2010, Valtonen2010a, Valtonen2016a} could accurately estimate the spin of the primary SMBH and show its significance in the orbital evolution. Pihajoki \cite{Pihajoki} showed that the secondary SMBH  could also produce bremsstrahlung outbursts, which implies that the auxiliary SMBH can also spin up. We found that in the analysis of Gravitational Waves (GW), the contribution of the secondary spin is not significant enough to be considered in this article. Here we used the highest spin--spin\cite{FBB, FBB2}, spin--orbit\cite{BoheSOa, BoheSOb} terms and post-Newtonian \cite{CBwaves, Owen, MoraWill, Kidder, IyerWill, BlanchetLRR} (PN) up to the newly released 4th PN order\cite{4PN}.

	OJ 287 is composed of a pair of close SMBHs, and such a binary system emits intense gravitational radiation. The large amplitude of its radiation makes OJ 287 an excellent target for low-frequency gravitational-wave detectors\cite{Zhang2013}.  Earlier, Y. Zhang and his colleagues analyzed the gravitational waveform of OJ 287. They made their analysis up to 3.5PN order and their results for three periods are in harmony with the findings of Valtonen. Furthermore, they investigated the detectability of GWs emitted by the system\cite{Chen2018}. From the frequency of the binary system, it appears that we can detect them with Pulsar Timing Array\cite{PTA} (PTA) experiments. In this paper, we extend their resuts by going until the start of the merger phase, with the use of 4PN and spin terms. We will examine the orbital evolution and waveforms after the last stable orbit (LSO) and use OJ 287 as an experimental case for testing 4PN results.

	
%
%
%

\clearpage
	\section{Orbital evolution and spin effects}
	The inspiral of compact astrophysical objects can be described accurately in the weak-field limit, where the gravitational potential and orbital velocity are regarded as small parameters. In the post-Newtonian regime\cite{BlanchetLRR}, the motion of the binary system can be characterized as perturbed Keplerian orbits. The equation of motion and the emitted radiation of these systems are analyzed in detail by \cite{DDPLA, Kidder, IyerWill, GopaIyer, MoraWill, FBB, FBB2, WillPNSO, WangWill, Arun08}. As a result, thoroughly tested and ready to use formulas are available up to 4PN order.

	To study the orbital motion, the evolution of spins and angular momenta of the binary system, we used the computational tool called CBwaves \cite{CBwaves}. The equation of motion was integrated numerically by a fourth-order Runge--Kutta method. We determined the radiation field by a simultaneous evaluation of the analytic waveform involving all the contributions up to 2PN order. As an output of the code, both time and frequency domain waveforms are available. The PN contributions to the acceleration and the radiation field are listed in the appendices of \cite{CBwaves}.

	Improving previous results on spin effects, we added 3.5PN spin--orbit contribution to the equations of motion, see e.g. \cite{Owen, FBB}, and spin precession and the total-energy flux emitted in gravitational waves, as it can be found in \cite{BoheSOa, BoheSOb}. Furthermore, in 2018, Blanchet et al. \cite{4PN} released the fourth post-Newtonian term for acceleration in the center-of-mass frame. The differential equation of the motion solved by CBwaves can be written as the sum of the following terms

	\begin{align}\label{acceleration}
		\mathbf{\ddot{r}} =& \mathbf{\ddot{r}}_{PN} + \mathbf{\ddot{r}}_{2PN} + \mathbf{\ddot{r}}_{SO} + \mathbf{\ddot{r}}_{SS} + \mathbf{\ddot{r}}_{RR} + \mathbf{\ddot{r}}_{PNSO} + \mathbf{\ddot{r}}_{3PN} \nonumber\\
 		+& \mathbf{\ddot{r}}_{RR1PN} + \mathbf{\ddot{r}}_{3.5PNSO} + \mathbf{\ddot{r}}_{RRSO} + \mathbf{\ddot{r}}_{RRSS} + \mathbf{\ddot{r}}_{4PN} \ .
	\end{align}

	The exact form of the different contributions can be found in the appendix of \cite{CBwaves} up to 3PN. For the sake of completeness, the newly introduced 3.5PN spin-orbit terms, $\ddot{r}_{3.5PNSO}$, spin precession, and 4PN corrections are presented here in the Appendixes.

	We have obtained the next-to-leading order of spin-orbit term in the center-of-mass frame up to 3.5PN order from \cite{BoheSOa}, and the equation we used is Eq. $3.8$, which can be expressed as

	\begin{equation}
		\mathbf{\ddot{r}}_{\mathrm{3.5PNSO}} = \frac{Gm}{r^3}\left[ {}^{(3.5)}\mathbf{b}_1 +{}^{(3.5)}\mathbf{b}_2 \frac{G m }{r} +{}^{(3.5)}\mathbf{b}_3 \frac{G^2m^2}{r^2}\right],
	\end{equation}
	where the exact form of the coefficients $(\mathbf{b}_1,\mathbf{b}_2,\mathbf{b}_3)$ can be found in \cite{BoheSOa}. Also, we can find the corrections to the spin evolution in \cite{BoheSOa}, and in Appendix B, where we restrict ourselves to the following form
	\begin{equation}
		\mathbf{S}_1 =\mathbf{ n \times v } \left[
		\frac{1}{c^2}\alpha_\mathrm{1PN}
		+\frac{1}{c^4}\alpha_\mathrm{2PN}
		+\frac{1}{c^6}\alpha_\mathrm{3PN}
		+ \mathcal{O}\left(\frac{1}{c^7}\right)\right].
	\end{equation}


	Simultaneously with the orbital evolution, the radiation field $h_{ij}$ is determined by the analytic waveform contributions up to 2PN order in harmonic coordinates, valid for general eccentric and spinning sources. This can be decomposed as
    \begin{eqnarray}\label{eq:04}
        h_{ij} =& \frac{2G\mu}{c^4 D} [Q_{ij} + P_{0.5}Q_{ij} + PQ_{ij}+P^{1.5}Q_{ij}+P^2Q_{ij} + PQ_{ij}^{SO} \\  \nonumber
        &+ P^{1.5}Q_{ij}^{SO} + P^2Q_{ij}^{SO} + PQ_{ij}^{SS} + P^{1.5}Q_{ij}^{tail}] ~ , 
    \end{eqnarray}
    where $D$ is the distance to the source, and $\mu = m_1 m_2/(m_1 + m_2)$ is the reduced mass of the binary. It contains spin-orbit (SO), spin-spin (SS), and tail terms. The full expressions of these terms are found, in detail, in the appendix of \cite{CBwaves}. Using a fourth-order Runge-Kutta method, we could model the evolution and the emitted gravitational waves of general binary systems with high accuracy.

	 Looking at a plane wave traveling in direction $\mathbf{\hat{N}}$, which is the unit spatial vector pointing from the center of mass of the source to the observer, the transverse-traceless (TT) part of the radiation field is provided by
	\begin{equation}
		h_{ij}^{TT}= \Lambda_{ij,kl} h_{kl},
	\end{equation}
	where
	\begin{equation}
		\Lambda_{ij,kl}(\mathbf{\hat{N}}) = \mathbf{P}_{ik}\mathbf{P}_{jl} - \frac{1}{2}\mathbf{P}_{ij}\mathbf{P}_{kl}
	\end{equation}
	and
	\begin{equation}
		\mathbf{P}_{ij}(\mathbf{\hat{N}}) = \delta_{ij} - N_i N_j.
	\end{equation}
	
	The radiation field can be defined by choosing an orthonormal triad as in \cite{CBwaves}, where
	\begin{align}
		\mathbf{\hat{N}} &= (\sin \iota \cos \phi, \sin \iota \sin \phi, \cos \iota), \\
		\mathbf{\hat{p}} &= (\cos \iota \cos \phi, \cos \iota \sin \phi, -\sin \iota), \\
		\mathbf{\hat{q}} &= (-\sin \phi, \cos \phi, 0),
	\end{align}
	  and $\iota$ and $\phi$ are the polar angles determining the relative position of the radiation frame with respect to the source frame $(\mathbf{\hat{x}},\mathbf{\hat{y}},\mathbf{\hat{z}})$, see the first figure of \cite{CBwaves}. The polarization states of the gravitation wave can be given, with respect to the orthonormal radiation frame, as
	\begin{equation}
		h_+ = \frac{1}{2}\left(\hat{p}_i\hat{p}_j - \hat{q}_i\hat{q}_j \right)h_{ij}^{TT}, \quad h_\times = \left( \hat{p}_i\hat{q}_j - \hat{q}_i\hat{p}_j \right)h_{ij}^{TT}.
	\end{equation}
	The strain produced by the binary system at the detector can be given with the combination
	\begin{equation}
		h(t) = F_+ h_+(t) + F_\times h_\times(t),
	\end{equation}
	where $F_+$ and $F_\times$ are the antenna pattern functions, which can be found in detailed form in \cite{CBwaves}.
	\pagebreak
	\section{Numerical results}
	In our numerical analysis, we used the software package CBwaces developed by P. Csizmadia \cite{CBwaves} and maintained by our group. This package provides a fast and accurate computational tool to examine gravitational waves emitted by spinning eccentric binary systems. We achieve this by integrating the center-of-mass equation of motion within the post-Newtonian framework up to 4PN order, and the spin precession equations up to 3PN developed by Bohé et al. \cite{BoheSOa, BoheSOb}.
    
   The initial parameters for our analysis are chosen from the estimates of Voltanen \cite{Valtonen2010, Valtonen2010a}, and Y. Zhang \cite{Zhang2013}. The values are listed in Table \ref{table1}.
    
	\begin{table}[h]
		\centering
		\caption{Initial parameters}
		{ \begin{tabular}{rc|r}
			$ m_\odot $ & (meter) & 1476.62504 \\
			$ m_1 $ & $(M_\odot) $ & $ 1.84 \times 10^{10} $ \\
			$ m_2 $ & $ (M_\odot) $ &  $ 1.46 \times 10^8 $ \\
			$ r_0 $ & $ (M) $ & $ \sim 100 $ \\
			$ S_1/m_1^2 $& & $ 0.381 $ \\
			$ S_2/m_2^2 $& & $0.996$  \\
			$\Delta \phi$& $({\deg})$ & $39.1$ \\
			$e_0$ & & $0.658 $ \\
		\end{tabular}\label{table1} }
	\end{table}
	Here $m_1$ is the mass of the primary SMBH and $m_2$ is the mass of the secondary SMBH, $r_0$ is the initial separation which we set to be approximately $100M$ with $(M=m_1+m_2)$, $\Delta\phi$ is the precession rate of the orbit per period, $e_0$ is the initial eccentricity of the auxiliary mass' orbit. We have chosen a different initial separation from the value, $r_0 \sim 0.535~\text{pc} \simeq 60~\text{M}$, which can be found in \cite{Zhang2013} for the sake of simplicity.

\begin{figure}
		\centering
		\includegraphics[width=0.7\textwidth]{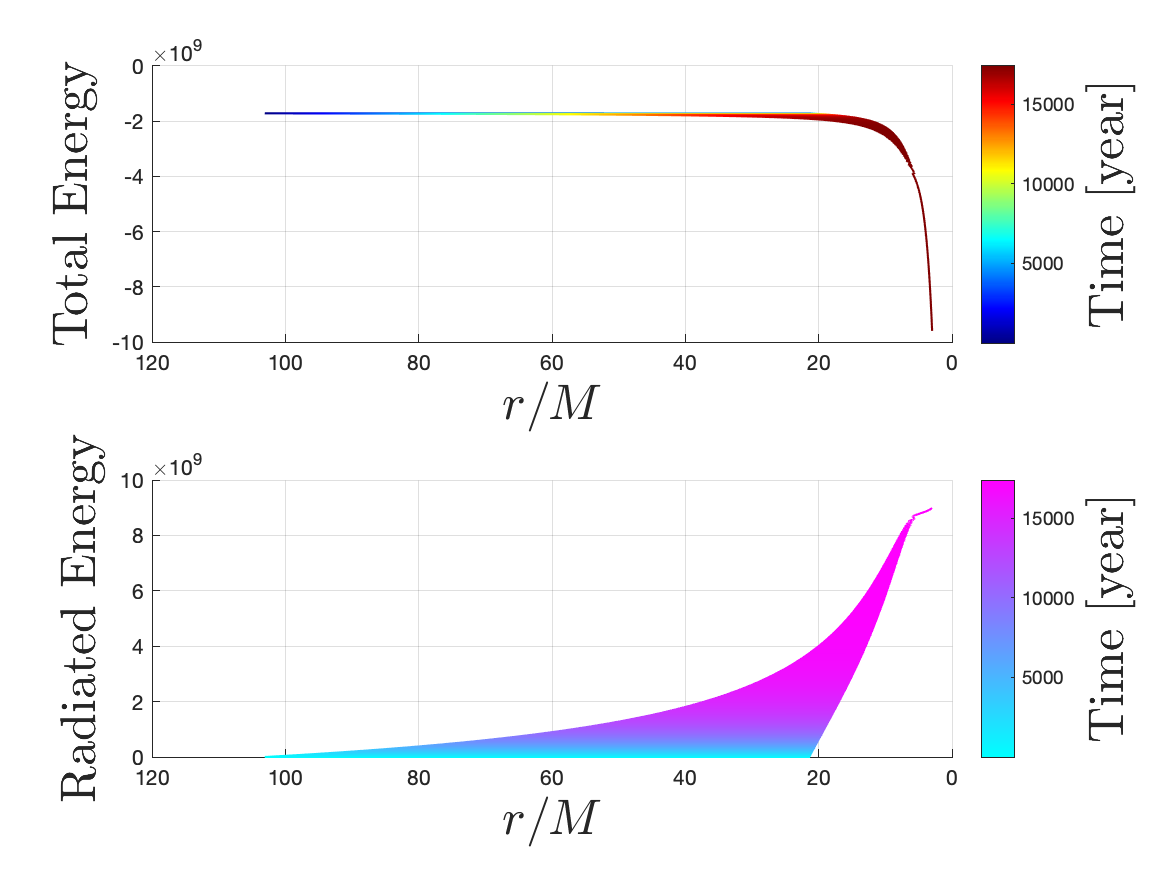}
		\caption{The upper panel of the figure, we can see the 4PN-accurate orbits of the secondary SMBH, whereas the central and lower panels show the total and radiated energies of the binary system in geometric units ($G=c=1$). Before the separation at 3M, the orbits computed in our 4PN-accurate model are in good agreement with the literature and do not expose any peculiar characteristics due to numerical errors. The total and radiated energies, respectively, are in line with our prior expectations.}\label{fig:03}
	\end{figure}

	In \cite{Pihajoki}, we found that both SMBH owns spin, but in our analysis, we assumed that only the primary SMBH has a spin, and it points to the positive direction, perpendicular to the orbital plane. In earlier computations, we have found that the contribution of the secondary SMBH's rotation is relatively small in the waveform and the orbital evolution of OJ 287. We had run simulations of the binary system in the inspiral phase and let the separation of SMBHs evolve beyond the last stable orbit, or into the so-called unstable orbits. We have estimated that the merger of the two SMBH would start slightly later than $\sim 17000~\text{years}$. We stopped our simulation before the final unstable circular orbit ends, in other words, before the separation of the two bodies becomes smaller than $r=3M$. With this requirement, we also prevented the event horizons from coming to contact with each other, which would be beyond the capabilities of the program package we used. The orbits of the secondary SMBH and the total and radiated energies have been presented in Figure \ref{fig:03}. As we can see, as long as the two SMBH are separated more than $6M$, the total energy of the system remains conserved. But by looking at the radiated energy we see it almost stops growing as the auxiliary SMBH converges in. It gives us some confidence that the calculations still give physically relevant results even beyond the LSO.
	
	\begin{figure*}[h!]
		\centering
		\includegraphics[width=0.7\textwidth]{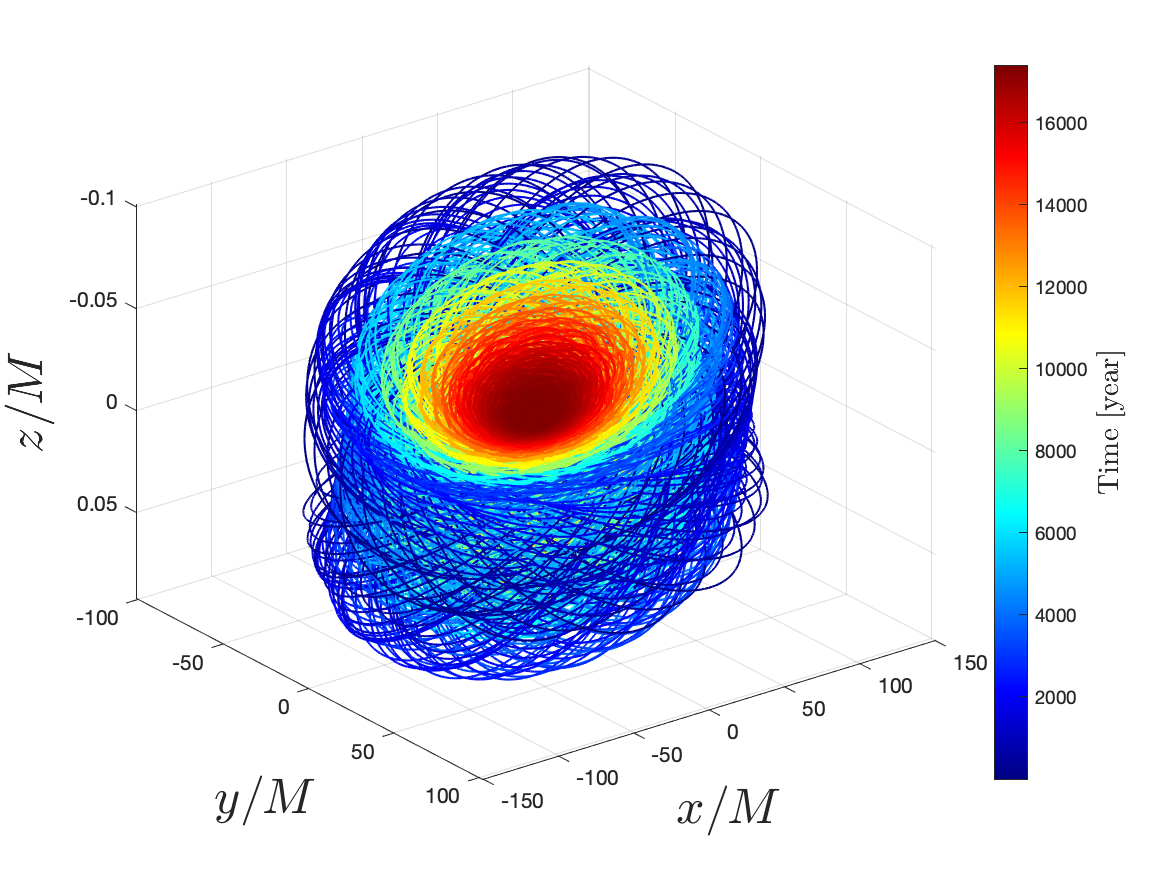}
	\end{figure*}

\begin{figure}[ht]
		\centering
		\includegraphics[width=.6\textwidth]{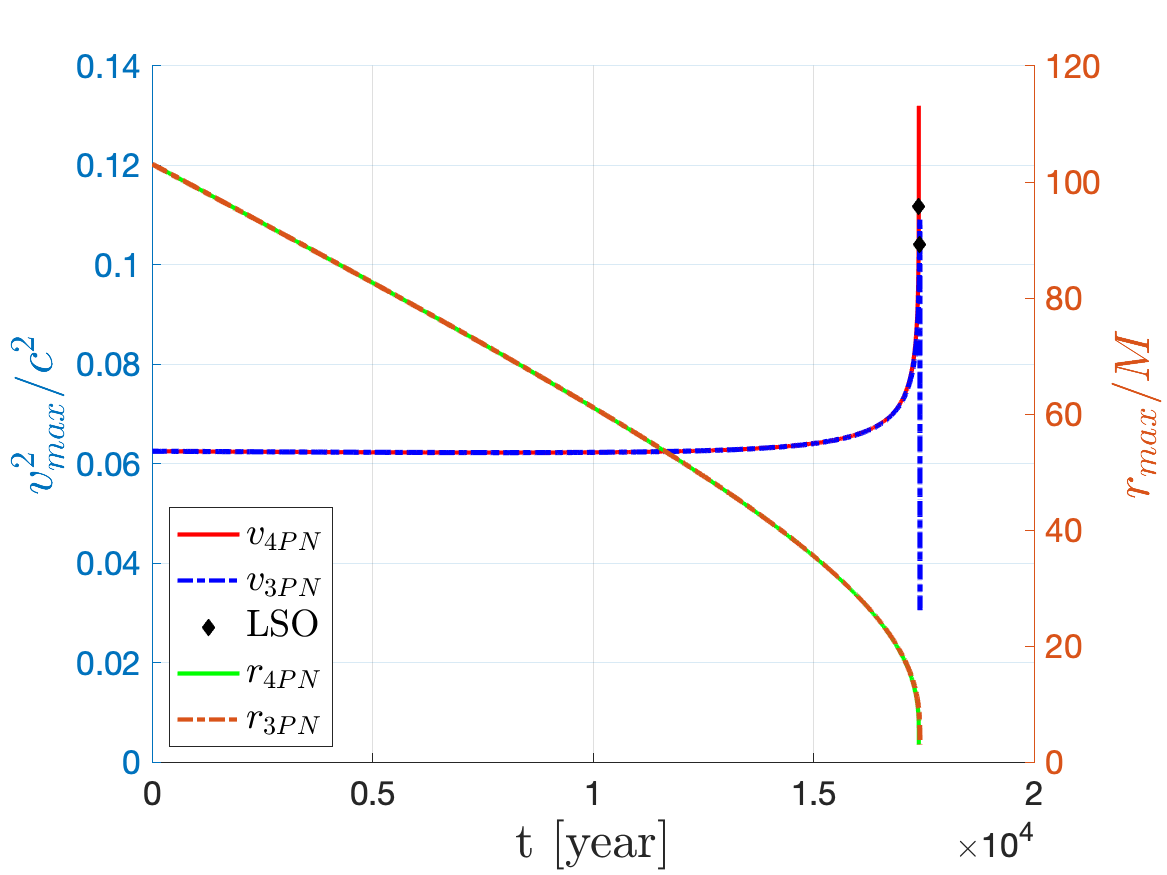}
		\caption{The maximum mutual separation $r_{max}$ (right $y$ axis) shows that in the early phase of the inspiral 4PN contributions are small, while the maximum speed of the binary drops for 3PN near the 3M limit.}\label{fig:04}
	\end{figure}

	In order to support our hypothesis, we have analyzed the maximum separation and post-Newtonian parameter $\epsilon \sim (v/c)^2 \sim GM/(rc^2)$ of the 3PN- and 4PN-accurate simulations, as shown in Figure \ref{fig:04}.

	As we can see in Figure \ref{fig:04}, the separation in both 3PN and 4PN approximations, marked by solid green and dashed orange lines, changes similarly. On the other hand, the post-Newtonian parameters in the 3PN and 4PN approximations are different beyond the LSO. The 3PN order's post-Newtonian parameter starts to decrease shortly after leaving the LSO, while at 4PN order it grows steadily. Consequently, the 4PN contribution provides us with a more accurate description of the orbital evolution beyond the LSO than the computations with the 3PN-accurate equation.  According to reference \cite{CBwaves}, PN formalism is valid if $\epsilon \sim 0.08 - 0.1$. However, in Section 9.6 of reference \cite{BlanchetLRR}, Blanchet argued that at LSO orbital velocities are of the order of 50\% of the speed of light. During our computations, the orbital velocity was around $v = \sqrt{\epsilon} \sim 0.35*c$,  which is below this limit.

\begin{figure}
		\centering
		\includegraphics[width=0.6\textwidth]{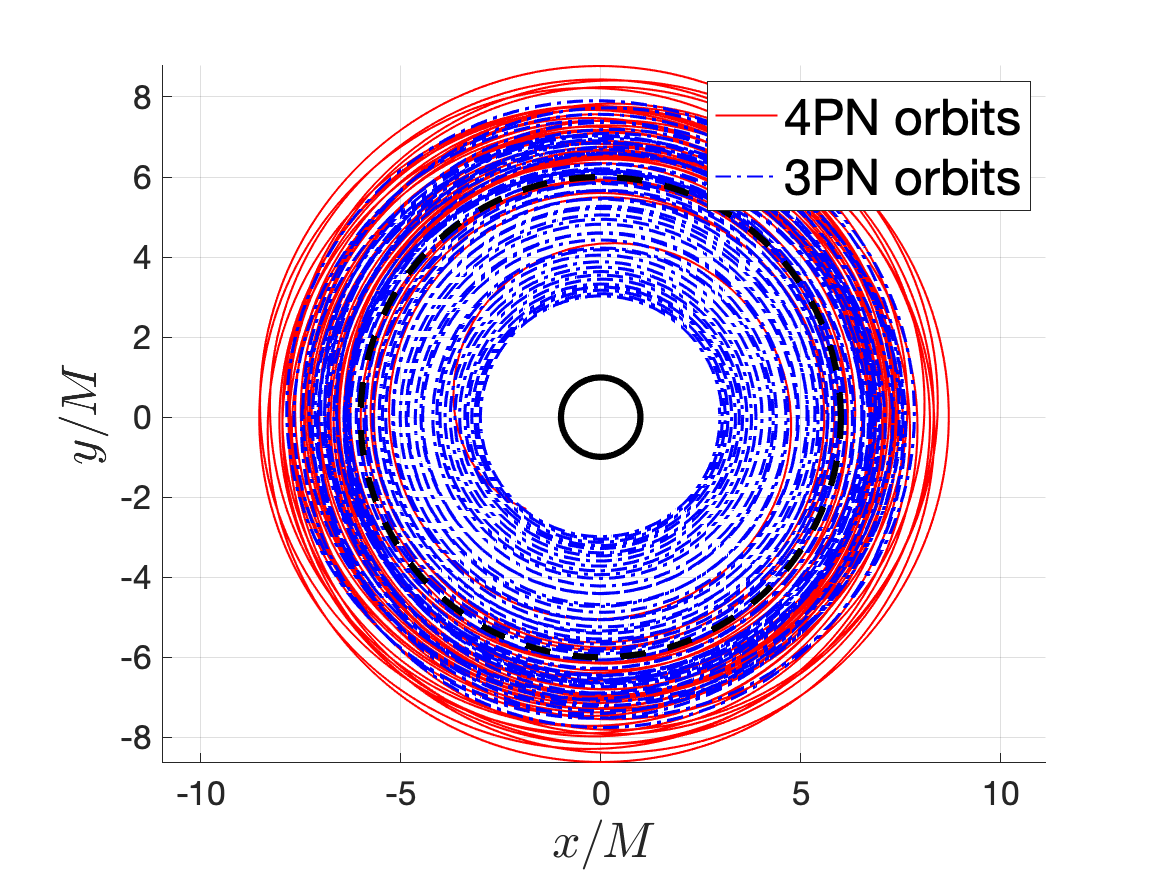}
		\caption{The unstable circular orbits of the secondary SMBH in 4PN and 3PN approximations are marked by red and blue lines, respectively. Solid and dashed black circles represent the event horizon and the LSO, respectively.} \label{fig:05}
	\end{figure}

\begin{figure}
		\centering
		\includegraphics[width=0.6\textwidth]{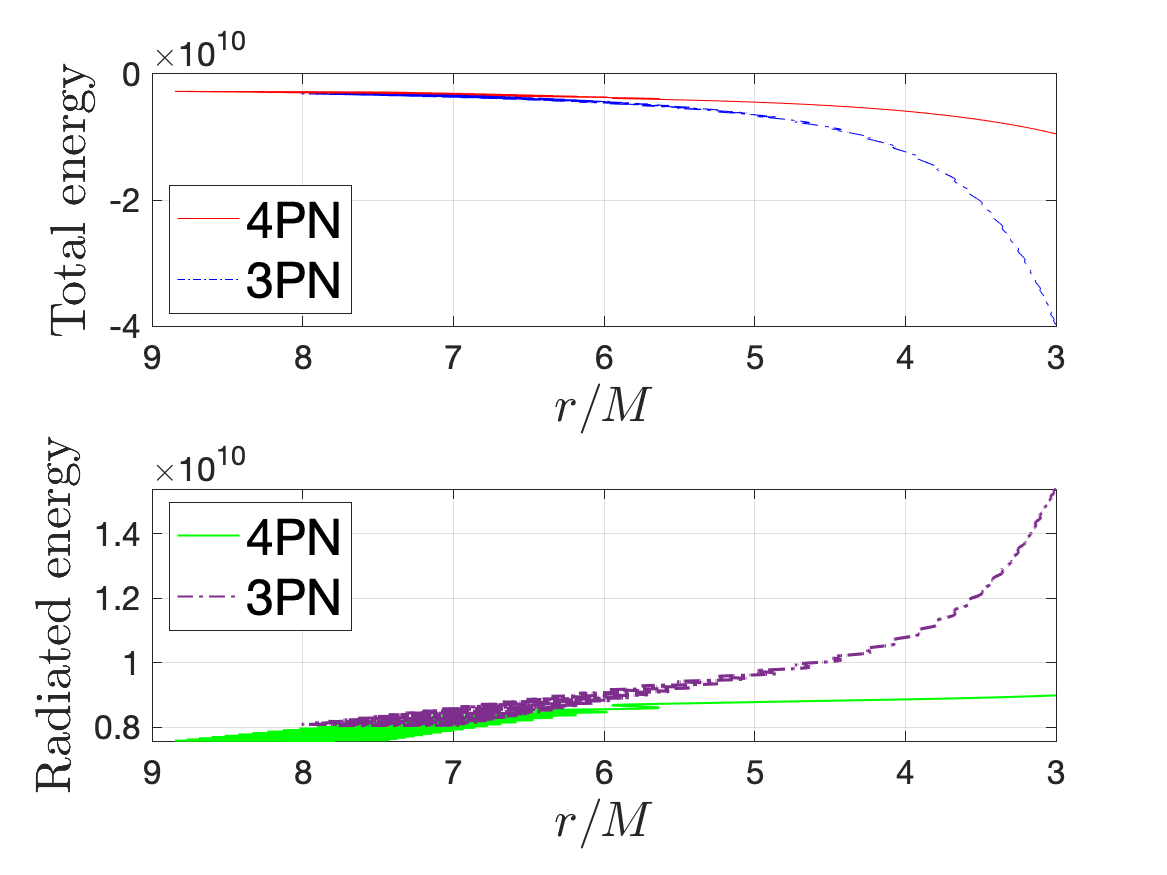}
		\caption{The total and the radiated energies of the unstable orbits of OJ 287 are presented in this figure. The energies are in geometric units ($G=c=1$). This figure shows that 4PN contributions make the equation of motion numerically more accurate than stopping at 3PN. It means that we require 4PN terms to analyze unstable orbits of eccentric, extreme mass ratio, and supermassive BH binary systems.} \label{fig:06}
	\end{figure}

	Following the computations presented above, now we can analyze the unstable orbits in the 4PN regime. We have shown the orbital evolution in 3PN and 4PN accuracy in Figure \ref{fig:05}. The total and radiated energies of the system are shown in Figure \ref{fig:06}.

	
	Figure \ref{fig:05} and \ref{fig:06} show further proof of our assumption, as we find more differences in the behavior orbits at 3PN and 4PN orders. We notice that the orbits at 3PN order make a similar number of cycles before and after the LSO. The orbits at 4PN order reach the 3M limit within a few revolutions beyond the LSO. Furthermore, if we look at Figure \ref{fig:06}, we see that the total and radiated energies of the binary start to diverge earlier in 3PN order. Following the arguments presented above, we can declare that the analysis of unstable orbits of eccentric, extreme mass ratio, and supermassive BH binary systems require the 4PN accurate description of orbital motion. \pagebreak

\begin{figure}
		\centering
		\includegraphics[width=.6\textwidth]{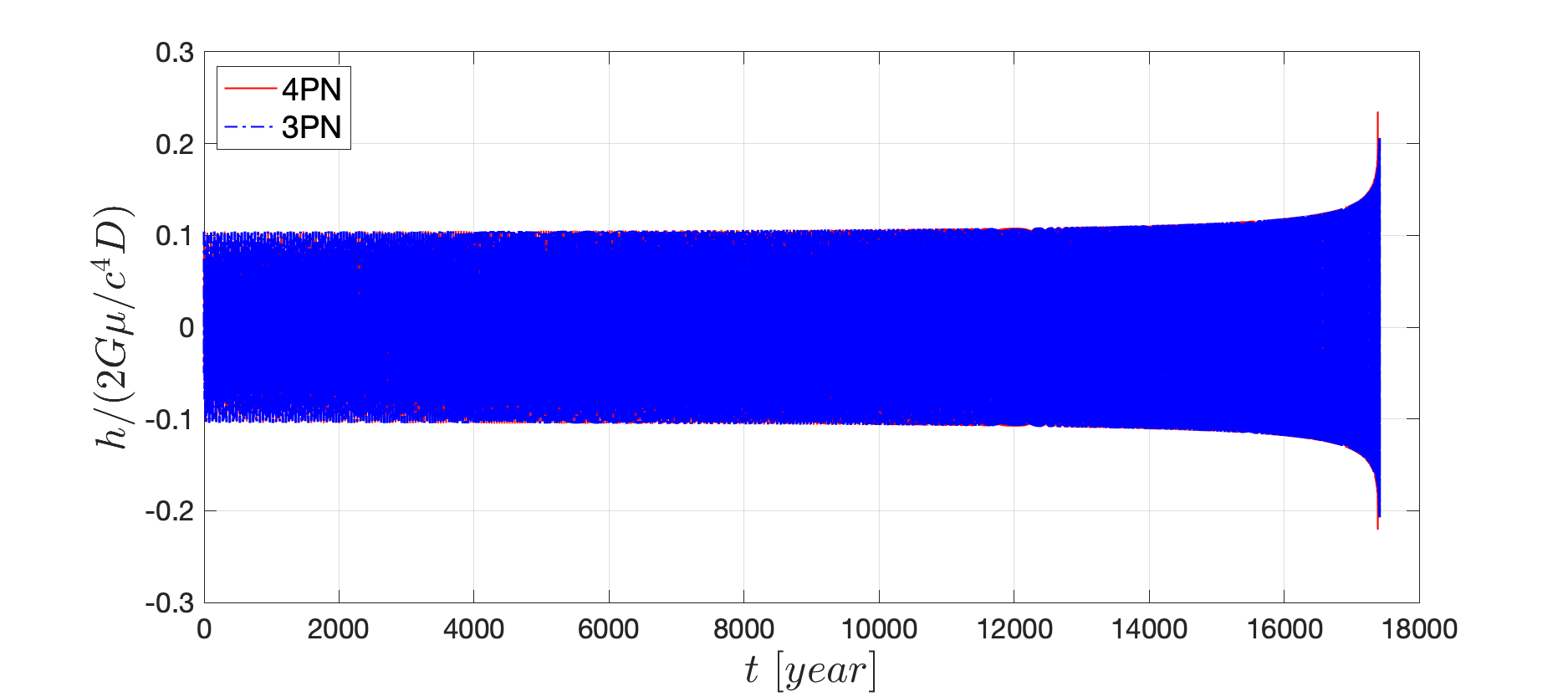}
		\caption{Gravitational waves emitted by the binary during its complete orbital evolution. We compared the waveforms provided by 3PN and 4PN terms.}\label{fig:07}
	\end{figure}

\begin{figure}
		\centering
		\includegraphics[width=.6\textwidth]{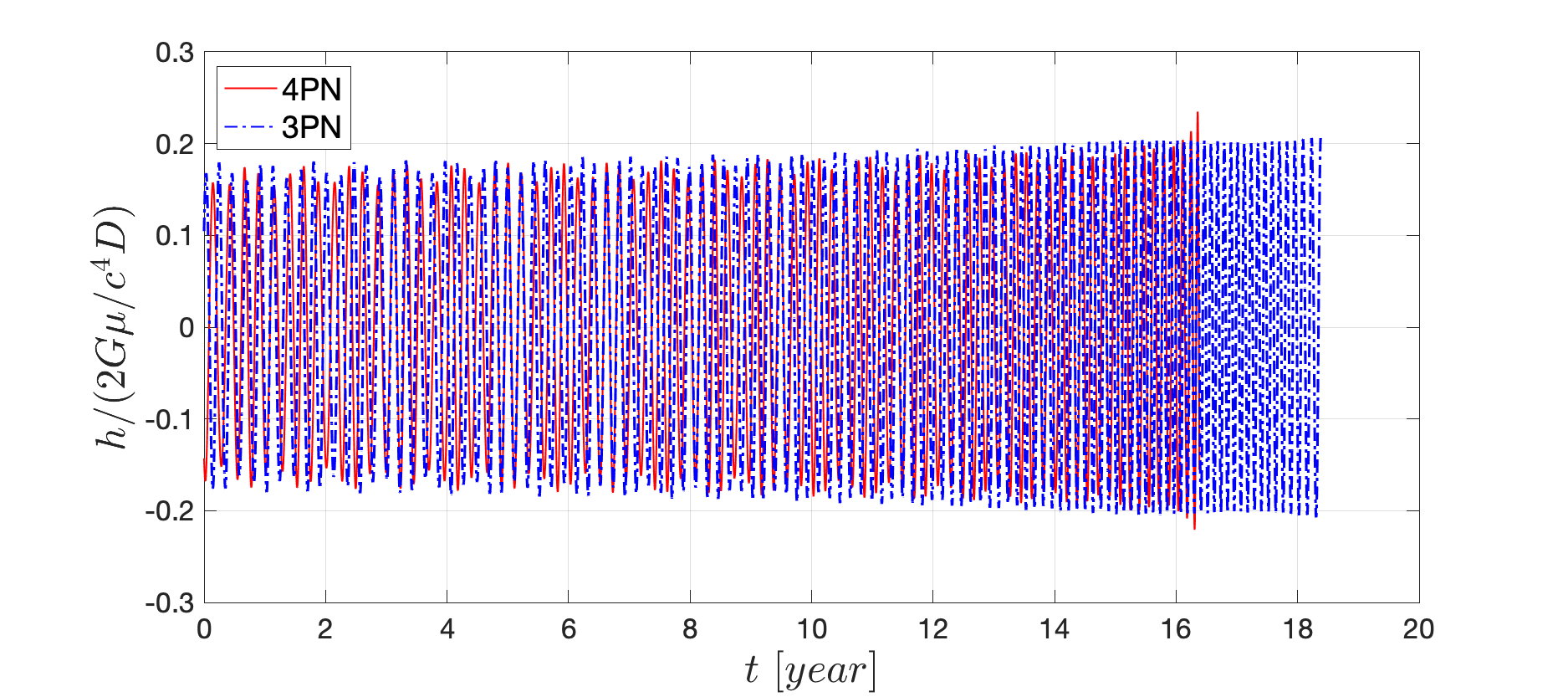}
		\caption{Gravitational waves strains emitted beyond the LSO.}\label{fig:08}
	\end{figure}

	We have shown in Figure \ref{fig:07} and \ref{fig:08} the strain of the gravitational waves emitted by the binary, for the entire orbital evolution, and for orbits after the LSO. In order to obtain the proper amplitudes, one has to know the distance of the source from the observer. We used the distance $D$ found in \cite{Zhang2013} as follows
	\begin{equation}
		D_L(z) = \frac{c}{H_0}(1+z) \int \limits_0^z \frac{dz'}{\sqrt{\Omega_m(1+z')+\Omega_\Lambda}},
	\end{equation}
	where  $\Omega_m = 0.685$, $\Omega_\Lambda = 1 - \Omega_m$ are cosmological constants in a spatially flat $\Lambda$CDM universe, and $H_0 = 67.3~\text{(km/s)/Mpc}$ is the Hubble constant. Then the distance of OJ 287 is $D = D_L(0.306) = 1.64689\times 10^9$ pc. Consequently, the prefactor amplitude of equation \ref{eq:04} is $\frac{2G\mu}{c^4D} = 8.6539 \times 10^{-15}$.


\begin{figure}
		\centering
		\includegraphics[width=0.7\textwidth]{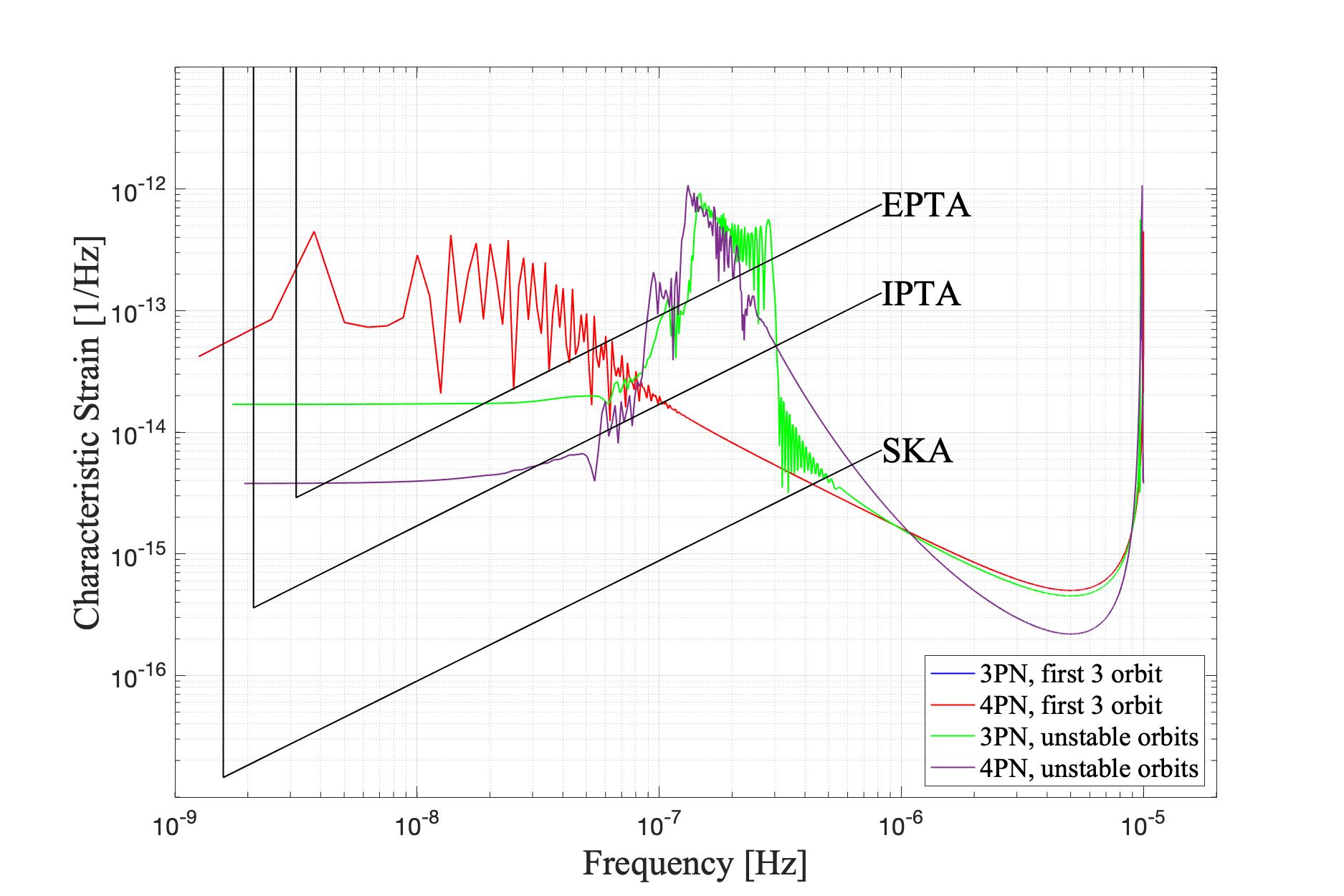}
		\caption{In this figure we have shown the characteristic strain of OJ 287. Red and blue lines represent the strain for 4PN and 3PN initial orbits. Similarly, purple and green lines represent the strain for unstable orbits. Furthermore, we provided the detector sensitivity curves for EPTA, IPTA, and SKA experiments with black lines. Let us note that, for the initial orbits, the contribution of 4PN terms is so small that 3PN and 4PN accurate strains cannot be individually distinguished.}\label{fig:09}
	\end{figure}

	As we have shown above, the amplitude remains $h \sim 10^{-15}$ while the frequency of the binary drops to $f\simeq 6.85 \times 10^{-8}$ Hz. It shows that, with the current interferometers on Earth and space telescopes like LISA, even this close to each other, OJ 287 is not a feasible candidate for detection. On the other hand, OJ 287 is an excellent target for Pulsar Timing Array experiments like EPTA, IPTA, and SKA. The sensitivity curves of these detectors and the characteristic strain of OJ 287 are presented in Figure \ref{fig:09}. The characteristic strain is attained by Fourier transformation of the gravitational wave amplitude,
	\begin{equation}
		\tilde{h} = \mathcal{F}\{h(t)\}(f) = \int \limits_{-\infty}^\infty h(t)e^{-2\pi ift}~dt,
	\end{equation}
	as it is found in reference \cite{Moore2015}. Then we calculated the strain for the first few orbits of our simulation and the unstable orbits in 3PN and 4PN orders. As we have seen earlier, the contribution of 4PN terms to the equation of motion is initially small ($\Delta a_{3PN}\sim 10^{-32}$ and $\Delta a_{4PN}\sim 10^{-35}$) compared to 3PN.
	
\newpage
	\section{Conclusions}
	In this paper, we have studied the orbital evolution of the supermassive black hole binary system OJ 287. We applied the post-Newtonian \cite{BlanchetLRR} approximation up to the fourth-order correction \cite{4PN} term and used the next-to-next-leading order of spin effects \cite{BoheSOa, BoheSOb, FBB, FBB2}. We have examined the evolution of the system beyond the last stable orbit (LSO). By explicit computations of CBwaves, we have found that the 4PN order is essential in the evolution of orbits beyond the LSO since it makes a more numerically accurate evolution than the 3PN order. We have proved that by analyzing the total and radiated energies and the post-Newtonian parameter of the system in detail, see fig. \ref{fig:04} and \ref{fig:06}. We have shown the waveform for the entire inspiral phase and presented the difference between the waveforms for 4PN and 3PN orders, see fig. \ref{fig:08}.

	Unfortunately, because of its low frequency, we cannot detect this binary with our detectors in the solar system, even though OJ 287 is a perfect candidate for detection techniques that target astrophysical objects like neutron stars, as we can see in the figure \ref{fig:09}. These are the pulsar timing\cite{PTA} experiments, including European Pulsar Timing Array\cite{EPTA} (EPTA), International Pulsar Timing Array\cite{IPTA} (IPTA) and Square Kilometer Array\cite{SKA} (SKA).	
	
	\section{Acknowledgments}

	This work was supported by COST Action PHAROS (CA16214), the Hungarian NKFIH Grants No. 124366 and 124508. Furthermore, we would like to acknowledge the support of Wigner GPU Laboratory.

\appendix

	\addtocontents{toc}{\protect\contentsline{chapter}{Appendix:}{}}
	\section{Center-of-mass 4th post-Newtonian order}\label{appendix:A}
	
	We can find the exact calculations of the 4th post-Newtonian order in \cite{4PN}, therefore, for the sake of simplicity, we can write the 4th order acceleration as
	\begin{equation}
	\mathbf{a}_{4PN} = -\frac{Gm	}{r^2}(A\mathbf{b} + B\mathbf{v})
	\end{equation}
	where we get the $A$ coefficient as the sum of the following terms
	\begin{align}
		A^{(0)} &= \left( \frac{315}{128}\eta - \frac{2205}{128}\eta^2 + \frac{2205}{64}\eta^3 - \frac{2205}{128}\eta^4 \right)\dot{r}^8 + \left( - \frac{175}{16}\eta + \frac{595}{8}\eta^2 - \frac{2415}{16}\eta^3 \right. \nonumber \\
		&+ \left. \frac{735}{8}\eta^4 \right)\dot{r}^6v^2 + \left( \frac{135}{8}\eta - \frac{1875}{16}\eta^2 + \frac{4035}{16}\eta^3 - \frac{1335}{8}\eta^4 \right)\dot{r}^4 v^4 + \left( -\frac{21}{2}\eta \right. \nonumber \\ 
		&+ \frac{1191}{16}\eta^2 - \left. \frac{327}{2}\eta^3 + 99\eta^4 \right)\dot{r}^2 v^6 + \left( \frac{21}{8}\eta - \frac{175}{8}\eta^2 + 61\eta^3 - 54\eta^4 \right)v^8 \\
		A^{(1)} &= \frac{m}{r} \left[ \left( \frac{2973}{40}\eta + 407\eta^2 + \frac{181}{2}\eta^3 - 86\eta^4 \right) \dot{r}^6 + \left( \frac{1497}{32}\eta - \frac{1627}{2}\eta^2 \right. \right. \nonumber \\
		&- 81\eta^3 + 228\eta^4 \bigg) \dot{r}^4 v^2 + \left( -\frac{2583}{16}\eta + \frac{1009}{2}\eta^2 + 47\eta^3 - 104\eta^4 \right)\dot{r}^2 v^4 \nonumber \\
		&+ \left. \left( \frac{1067}{32}\eta - 58\eta^2 - 44\eta^3 + 58\epsilon^4 \right)v^6 \right] \\
		%
		A^{(2)} &= \frac{m^2}{r^2} \left\{ \left[ \left( \frac{2094751}{960} + \frac{45255}{1024} \pi^2 \right)\eta + \frac{326101}{96}\eta^2 - \frac{4305}{128}\eta^3 - \frac{1959}{32}\eta^4 \right]\dot{r}^4  \right. \nonumber \\
		&+ \left[ -\left( \frac{1636681}{1120} + \frac{12585}{512} \pi^2  \right)\eta - \left(\frac{255461}{112}  + \frac{3075}{128}\pi^2 \right)\eta^2 - \frac{309}{4} \eta^3 + 63\eta^4 \right] \dot{r}^2 v^2 \nonumber \\
		&+ \left[ \left( \frac{1096941}{11200} + \frac{1155}{1024}\pi^2 \right) \eta + \left( \frac{7263}{70} - \frac{123}{64}\pi^2 \right)\eta^2 + \frac{145}{2}\eta^3 - 16\eta^4 \right] v^4 \\
		A^{(3)} &= \frac{m^3}{r^3} \left\{ \left[ -2 + \left( \frac{1297943}{8400} - \frac{2969}{16} \pi^2 \right)\eta + \left( \frac{1255151}{840} + \frac{7095}{32}\pi^2 \right)\eta^2 - 17\eta^3  \right. \right. \nonumber \\
		&- \left. 24\eta^4 \bigg]\dot{r}^2 \left[ \left( \frac{1237279}{252000} + \frac{3835}{96}\pi^2 \right)\eta - \left( \frac{693947}{2520} + \frac{229}{8}\pi^2 \right)\eta^2 + \frac{19}{2} \eta^3 \right]v^2  \right\} \\
		A^{(4)} &= \frac{m^4}{r^4} \left[ 25 + \left( \frac{6625537}{12600} - \frac{4543}{93} \pi^2 \right)\eta + \left( \frac{477763}{420} + \frac{3}{4}\pi^2 \right) \eta^2 \right]
	\end{align}
	and we get $B$ coefficient similarly to $A$ as:
	\begin{align}
		B^{(0)} &= \left( \frac{105}{16}\eta - \frac{245}{8}\eta^2 + \frac{385}{16}\eta^3 + \frac{35}{8}\eta^4 \right)\dot{r}^7 + \left( -\frac{165}{8}\eta + \frac{1665}{16}\eta^2 - \frac{1725}{16}\eta^3 - \frac{105}{4} \eta^4 \right)\dot{r}^5 v^2 \nonumber \displaybreak[3]\\
		&+ \left(\frac{45}{2}\eta - \frac{1869}{16}\eta^2 + 129\eta^3 + 54\eta^4 \right)\dot{r}^3 v^4 + \left( -\frac{157}{16}\eta + 54\eta^2 - 69\eta^3 - 24\eta^4 \right)\dot{r}v^6  \\
		B^{(1)} &= \frac{m}{r} \left[ \left( -\frac{54319}{160}\eta - \frac{901}{8}\eta^2 + 60\eta^3 30\eta^4 \right)\dot{r}^5 + \left( \frac{25943}{48}\eta + \frac{1199}{12}\eta^2 - \frac{349}{2}\eta^3 - 98\eta^4 \right)\dot{r}^3v^2 \right. \nonumber \\
		&+ \left. \left( - \frac{5725}{32}\eta - \frac{389}{8}\eta^2 + 118 \eta^3 + 44\eta^4 \right)\dot(r)v^4 \right] \\
		B^{(2)} &= \frac{m^2}{r^2} \left\{ \left[ - \left( \frac{9130111}{3360} + \frac{4695}{256}\pi^2 \right)\eta - \left( \frac{184613}{112} - \frac{1845}{64}\pi^2 \right)\eta^2 + \frac{209}{2}\eta^3 + 74\eta^4 \right]\dot{r}^3 \right. \nonumber \\
		&+ \left[ \left( \frac{8692601}{5600} + \frac{1455}{256}\pi^2 \right)\eta + \left( \frac{58557}{70} - \frac{123}{8}\pi^2 \right)\eta^2 -70\eta^3 - 34\eta^4 \right]\dot{r}v^2 \\
		B^{(3)} &= \frac{m^3}{r^3} \left[ 2 - \left( \frac{619267}{525} - \frac{791}{16}\pi^2 \right)\eta - \left( \frac{28406}{45} + \frac{2201}{32}\pi^2 \right)\eta^2 + 66\eta^3 16\eta^4 \right]\dot{r} 
	\end{align}
	where $\eta = \mu/(m_1 + m_2)$ the reduced mass, $\mu = m_1 m_2/(m_1 + m_2)$, $m = m_1 + m_2$ the total mass of the binary system and $\mathbf{n} = \mathbf{r}/r$ the normal vector. Furthermore $v^2$ is the absolute value of the speed of the center-of-mass, which we derive from $v^2 = \mathbf{v \cdot v}$, where $\mathbf{v} = \mathbf{v_1} - \mathbf{v}_2$ is the relative velocity. At last $\dot{r} = \mathbf{n \cdot v}$ is the scalar product of the realtive velocity and the normal vector.
	
	Now we can turn our attention to the 4th order corrections of the $E_{4PN}=E/\mu$ energy and $\mathcal{L} = \mathcal{L}/\mathcal{L}_N$ orbital angular momentum of the system. First let see the terms of the energy corrections as it can be found in the 3rd section of \cite{4PN}
	\begin{align}
		\mathcal{E}^{(0)} &= \left( \frac{63}{256} - \frac{1089}{256}\eta + \frac{7065}{256}\eta^2 - \frac{10143}{128}\eta^3 + \frac{21735}{256}\eta^4 \right)v^{10} \\
		\mathcal{E}^{(1)} &= \frac{m}{r} \left[ \left( -\frac{35}{128}\eta + \frac{245}{128}\eta^2 - \frac{245}{64}\eta^3 + \frac{245}{128}\eta^4 \right)\dot{r}^8 + \left( \frac{25}{32}\eta - \frac{125}{16}\eta^2 + \frac{185}{8}\eta^3 - \frac{595}{32}\eta^4 \right)\dot{r}^6 v^2 \right. \nonumber \\
		&+ \left( \frac{27}{64}\eta + \frac{243}{32}\eta^2 - \frac{1683}{32}\eta^3 + \frac{4851}{64}\eta^4 \right)\dot{r}^4 v^4 + \left( - \frac{147}{32}\eta + \frac{369}{32}\eta^2 + \frac{423}{8}\eta^3 - \frac{4655}{32}\eta^4 \right)\dot{r}^2 v^6   \nonumber \\
		&+ \left. \left( \frac{525}{128} - \frac{4011}{128}\eta + \frac{9507}{128}\eta^2 - \frac{357}{64}\eta^3 - \frac{15827}{128}\eta^4 \right)v^8 \right] \\
		\mathcal{E}^{(2)} &= \frac{m^2}{r^2} \left[ \left( -\frac{4771}{640}\eta - \frac{461}{8}\eta^2 -\frac{17}{2}\eta^3 + \frac{15}{2}\eta^4 \right)\dot{r}^6 + \left( \frac{5347}{384}\eta + \frac{19465}{96}\eta^2 - \frac{439}{8}\eta^3 \right. \right. \nonumber \\
		&- \left. \frac{135}{2}\eta^4\right)\dot{r}^4v^2 + \left( \frac{15}{16} - \frac{5893}{128}\eta - \frac{12995}{64}\eta^2 + \frac{18511}{64}\eta^3 + \frac{2846}{16}\eta^4 \right)\dot{r}^2v^4 \nonumber \\
		&+ \left. \left( \frac{575}{32} - \frac{4489}{128}\eta + \frac{5129}{64}\eta^2 - \frac{8289}{64}\eta^3 + \frac{975}{16}\eta^4 \right)v^6 \right] \displaybreak[3]\\
		\mathcal{E}^{(3)} &= \frac{m^3}{r^3} \left\{ \left[ - \left( \frac{2599207}{6720} - \frac{6465}{1024}\pi^2 \right)\eta - \left(\frac{103205}{224} - \frac{615}{128}\pi^2 \right)\eta^2 + \frac{69}{32}\eta^3 + \frac{87}{4}\eta^4 \right] \dot{r}^4 \right. \nonumber \\
		&+ \left[ \frac{21}{4} \left( \frac{1086923}{1680} + \frac{333}{512}\pi^2 \right)\eta + \left( \frac{206013}{560} + \frac{123}{64}\pi^2 \right)\eta^2 - \frac{2437}{16}\eta^3 - \frac{141}{2}\eta^4 \right]\dot{r}^2v^2 \nonumber \\
		&+ \left. \left[ \frac{273}{16} - \left( \frac{22649399}{100800} - \frac{1071}{1024}\pi^2 \right)\eta + \left( \frac{521063}{10080} - \frac{205}{128}\pi^2 \right)\eta^2 + \frac{2373}{32}\eta^3 - \frac{45}{4}\eta^4 \right]v^4 \right\} \\
		\mathcal{E}^{(4)} &= \frac{m^4}{r^4} \left\{ \left[ \frac{9}{4} - \left( \frac{1622437}{12600} - \frac{2645}{96}\pi^2 \right)\eta - \left( \frac{289351}{2520} + \frac{1367}{32}\pi^2 \right)\eta^2 + \frac{213}{8}\eta^3 + \frac{15}{2}\eta^4 \right] \dot{r}^2 \right. \nonumber \\
		&+ \left. \left[ \frac{15}{16} + \left( \frac{1859363}{16800} - \frac{149}{32}\pi^2 \right) \eta + \left( \frac{22963}{5040} + \frac{311}{32}\pi^2 \right)\eta^2 - \frac{29}{8}\eta^3 + \frac{1}{2}\eta^4 \right] \right\} \\
		%
		\mathcal{E}^{(5)} &= \frac{m^5}{r^5} \left[ -\frac{3}{8} - \left( \frac{1697177}{25200} + \frac{105}{32}\pi^2 \right)\eta - \left( \frac{55111}{720} + 11\pi^2 \right)\eta^2 \right]
	\end{align}
	and the terms of the angular momentum, where $\mathcal{L}_N = \mu \mathbf{r} \times \mathbf{v}$ the Newtonian term
	\begin{align}
		\mathcal{L}^{(0)} &= \left( \frac{35}{128} - \frac{605}{128}\eta + \frac{3925}{128}\eta^2 - \frac{5635}{64}\eta^3 + \frac{12075}{128} \right) v^8 \\
		\mathcal{L}^{(1)} &= \frac{m}{r}\left[ \left( -\frac{5}{8}\eta + \frac{15}{8}\eta^2 + \frac{45}{16}\eta^3 - \frac{85}{16}\eta^4 \right)\dot{r}^6 + \left( 3\eta - \frac{45}{4}\eta^2 - \frac{135}{16}\eta^3 + \frac{693}{16}\eta^4 \right)\dot{r}^4v^2 \right. \nonumber \\
		&+\left( - \frac{53}{8}\eta + \frac{423}{16}\eta^2 + \frac{299}{16}\eta^3 - \frac{1995}{16}\eta^4 \right)\dot{r}^2 v^4 + \left( \frac{75}{16} - \frac{151}{4}\eta + \frac{1553}{16}\eta^2 - \frac{425}{16}\eta^3 \right. \nonumber \\
		&- \left. \left. \frac{2261}{16}\eta^4 \right)v^6 \right] \\
		\mathcal{L}^{(2)} &= \frac{m^2}{r^2} \left[ \left( \frac{14773}{320}\eta + \frac{3235}{48}\eta^2 - \frac{155}{4}\eta^3 - 27\eta^4 \right)\dot{r}^4 + \left( \frac{3}{4} - \frac{5551}{60}\eta - \frac{256}{3}\eta^2 + \frac{4459}{16}\eta^3 \right. \right. \nonumber \\
		+& \left. \left. \frac{569}{4}\eta^4 \right)\dot{r}^2v^2 + \left( \frac{345}{16} - \frac{65491}{960}\eta + \frac{12427}{96}\eta^2 - \frac{3845}{32}\eta^3 + \frac{585}{8}\eta^4 \right)v^4 \right] \\
		\mathcal{L}^{(3)} &= \frac{m^3}{r^3} \left\{ \left[ \frac{7}{2} + \left( \frac{7775977}{16800} + \frac{426}{256}\pi^2 \right)\eta + \frac{121449}{560}\eta^2 - \frac{1025}{8}\eta^3 - 47\eta^4 \right]\dot{r}^2 + \left[ \frac{91}{4} \right. \right. \nonumber \\
		&- \left. \left. \left( \frac{13576009}{50400} - \frac{469}{256}\pi^2 \right)\eta + \left( \frac{276433}{5040} - \frac{41}{16}\pi^2 \right)\eta^2 + \frac{637}{8}\eta^3 - 15 \eta^4 \right]v^2 \right\} \\
		\mathcal{L}^{(4)} &= \frac{m^4}{r^4} \left[ \frac{15}{8} + \left( \frac{3809041}{25200} - \frac{85}{8}\pi^2 \right)\eta - \left(\frac{20131}{420} - \frac{663}{32}\pi^2 \right)\eta^2 - \frac{15}{4}\eta^3 + \eta^4 \right]
	\end{align}

	\section{The post-Newtonian corrections of the spin} \label{appendix:B}

	As it is found in \cite{BoheSOa}, we can write the 3rd post-Newtonian order of the $\mathbf{S}_1$ spin vector in the center of mass frame, as follows
	\begin{equation}
		\mathbf{S}_1 = \mathbf{n \times v}\left[ \frac{1}{c^2} \alpha_{PN} + \frac{1}{c^4} \alpha_{2PN} + \frac{1}{c^6} \alpha_{3PN} + \mathcal{O}\left( \frac{1}{c^7} \right) \right],
	\end{equation}
	where the individual corrections can be written as

	\begin{align}
		\alpha_{PN} &= \frac{Gm}{r^2} \left( \frac{3}{4} + \frac{1}{2}\eta - \frac{3}{4} \frac{\delta m}{m} \right),  \\
		\alpha_{2PN} &= \frac{Gm}{r^2} \left[ \left( -\frac{3}{2}\eta + \frac{3}{4}\eta^2 - \frac{3}{2}\eta \frac{\delta m}{m} \right) \dot{r}^2 + \left( \frac{1}{16} + \frac{11}{8}\eta - \frac{3}{8}\eta^2 + \frac{\delta m}{m} \left( -\frac{1}{16} \right. \right. \right. \nonumber \\
		&+ \left. \left. \left. \frac{1}{2}\eta \right) \right)  v^2 \right] + \frac{G^2 m^2}{r^3} \left( -\frac{1}{4} - \frac{3}{8}\eta + \frac{1}{2} \eta^2 + \frac{\delta m}{m} \left( \frac{1}{4} - \frac{1}{8}\eta \right) \right), \\
		\alpha_{3PN} &= \frac{Gm}{r^2} \left[ \left( \frac{15}{8}\eta - \frac{195}{32}\eta^2 + \frac{15}{16}\eta^3 + \frac{\delta m}{m} \left( \frac{15}{8}\eta - \frac{75}{32}\eta^2 \right) \right) \dot{r}^4 + \bigg( -3\eta \right. \nonumber \\
		& + \left. \frac{291}{32}\eta^2 - \frac{45}{16}\eta^3 + \frac{\delta m}{m} \left( -3\eta + \frac{177}{32}\eta^2 \right) \right) \dot{r}^2 v^2 + \left( \frac{1}{32} + \frac{19}{16}\eta -\frac{31}{8}\eta^2 \right. \nonumber \\
		&+ \left. \left. \frac{17}{16}\eta^3 + \frac{\delta m}{m} \left( - \frac{1}{32} + \frac{3}{4}\eta - \frac{11}{8}\eta^2 \right) \right) v^4 \right] + \frac{G^2 m^2}{r^3} \left[ \left( \frac{1}{4} - \frac{525}{32}\eta - \frac{159}{16}\eta^2 \right. \right. \nonumber \\
		&+ \left. \frac{13}{4}\eta^3 + \frac{\delta m}{m} \left( - \frac{1}{4} - \frac{75}{32}\eta - \frac{87}{16}\eta^2 \right) \right) \dot{r}^2 + \left( \frac{3}{16} + \frac{27}{4}\eta + \frac{75}{32}\eta^2 - \frac{9}{8}\eta^3 \right. \nonumber \\
		&+ \left. \left. \frac{\delta m}{m} \left( -\frac{3}{16} + \frac{9}{8}\eta + \frac{35}{32}\eta^2 \right) \right) v^2 \right] + \frac{G^3 m^3}{r^4} \left( \frac{7}{16} - \frac{9}{4}\eta - \frac{9}{8}\eta^2 + \frac{1}{2}\eta^3 \right. \nonumber \\
		&+ \left. \frac{\delta m}{m} \left( -\frac{17}{16} - \frac{1}{8}\eta - \frac{1}{8}\eta^2 \right) \right).
	\end{align}

	Here $\delta m = m_1 -m_2$ is the mass difference. We can get the expressions for $\mathbf{S}_2$ with a simple $\delta m \to -\delta m$ exchange.

\newpage

\bibliographystyle{ws-ijmpd}
\bibliography{Oj287Art.bib}

\end{document}